# Weak decays of B$_c$ meson in a QCD potential model

K K Pathak[1], S Bhattacharya[2] and T Das[3]* 

[1]Department of Physics, Arya Vidyapeeth College, Guwahati 781016, India

[2]Impact Hub Institute for Competitive Examination, Dimapur 797112, India

[3]Department of Physics, Madhab Choudhury College, Barpeta 781301, India

**Abstract:** The leptonic and semileptonic decay properties of $B_c$ meson are studied in a QCD potential model. The CKM element for the leptonic decay is calculated by evaluating its decay constant. With the help of the Isgur–Wise function, the form factors and the CKM element for semileptonic decay are also calculated. The variation in four-momentum square in the two decays demands two different scales of the strong coupling constant $\alpha_s$. Thus, a two-scale picture is evident from the study. We compute the decay constant $f_p$ for leptonic decay and form factors for semileptonic decay with two different $\Lambda$ scales in a specific prescription where it is related to the strong coupling $\alpha_s$. Within the potential model, we find the scale for leptonic decay is $\Lambda \leqslant 0.150$ GeV and for semileptonic decay is $\Lambda \geqslant 0.410$ GeV.

## 1. Introduction

In the heavy quark dynamics, the decay properties of $B_c$ meson are given the special interest as it is the only heavy meson containing two different flavors and decays only through weak interaction [1–5]. The exclusive semileptonic decay processes of heavy mesons have generated a great interest not only in extracting the accurate values of Cabbibo–Kobayashi–Maskawa (CKM) matrix elements but also in testing different theoretical approaches to describe the internal structure of hadrons. The hadron colliders recently provide a platform to investigate the production and decay properties of $B_c$ meson with the running of the LHC with the luminosity of about $\mathcal{L} \sim 10^{33}$ cm$^{-2}s^{-1}$, and it is expected around $10^9$ $B_c$ events per year [6]. The advantage of semileptonic decay processes is that the effects of the strong interaction can be separated from the effects of the weak interaction into a set of Lorentz-invariant form factors, which provide the essential information of the strongly interacting quark and gluon structure inside hadrons. These form factors are not directly computed in the experiment. Thus, the theoretical problem associated with analyzing semileptonic decay processes is essential for calculating the weak form factors and is studied rigorously in the present scenario [7]. However, it is evident from the hadronic current of leptonic and semileptonic decay that the four-momentum square is not the same for both the decays, and thus, a two-scale picture is arised to study the two decays.

Here in this manuscript we compute two different $\Lambda$ scales which are related to the strong coupling constant $\alpha_s$ in a specific prescription and then calculate different form factors.

## 2. The momentum transfer in leptonic and semileptonic decay

In the weak decay of $B_c$ meson, the charged weak current operator $J^\mu$ which couples the W-boson according to the interaction Lagrangian [8] is

$$\unicode{0x141}_{int} = \frac{-g}{\sqrt{2}}(J^\mu W_\mu^+ + J^{\mu^+} W_\mu^-), \quad (1)$$

where

$$J^\mu = \sigma V_{q_1,q_2} J^\mu = \sigma \bar{u_i} \gamma^\mu \frac{1}{2}(1 - \gamma_5) V_{q_1,q_2} d_j. \quad (2)$$

Here $i$ and $j$ run over the three quark generations, so that the field operators annihilate, and hence, the amplitudes of the decay processes are proportional to the CKM element $V_{q_1,q_2}$.

*Corresponding author, E-mail: tapashimcc@gmail.com

The hadronic current cannot be easily evaluated, since the quarks in the hadrons are not free and nonperturbative strong interaction effect plays the prominent role in describing the physical states. The long-distance effects, present in the formation of the bound meson states in the hadronic interactions, are parameterized to form factors. These form factors are functions of the momentum transfer and polarization states of the hadrons.

For leptonic decays, the initial state is unpolarized and the momentum transfer is constant with $q^2 = m^2$, and hence, the form factor becomes a single constant $f_p$, known as the decay constant of the meson.

Mathematically, the amplitude for a leptonic decay is written as [8]

$$A(M_{Q\bar{q}} \to l^- \bar{v}) = -\frac{G_F}{\sqrt{2}} V_{qQ} L^\mu H_\mu,\quad (3)$$

where the leptonic current $L^\mu$ can be written in terms of the Dirac spinors $u_l$ and $v_v$

$$L^\mu = \bar{u}_l \gamma^\mu (1 - \gamma_5) v_v.\quad (4)$$

The hadronic current for leptonic decay is very simple, since the only four vectors available to be constructed with the leptonic current are $q^\mu$, i.e.,

$$H^\mu = \langle 0|\bar{q}\gamma^\mu(1-\gamma_5)Q|M = if_p q^\mu\rangle.\quad (5)$$

Here $f_p$ is parametrized to absorb all the strong interaction effects. Since the two initial quarks must annihilate, the matrix element is sensitive to $f_p$.

For semileptonic decay of a meson M into another meson X, the amplitude takes the form [9]

$$A(M_{Q\bar{q}} \to X_{q\bar{q}} l^- \bar{v}) = -\frac{G_F}{\sqrt{2}} V_{qQ} L^\mu H_\mu.\quad (6)$$

Here the hadronic current

$$H^\mu = \langle X|\bar{q}\gamma^\mu(1-\gamma_5)Q|M\rangle\quad (7)$$

is not calculated in a simple manner as is done in the leptonic decay, since $q^2$ is different for event to event. Thus, $H^\mu$ can be expressed in terms of different form factors, which isolate the effects of strong interactions on amplitude. Thus, in the two weak decays, the $q^2$ is not the same, and hence, a two-scale picture for the strong coupling constant is evident for any theoretical study.

## 3. The potential model

In the study of heavy-light mesons, a specific potential model with Cornell potential [10] as the ingredient has been found to be successful to some extent [11, 12]. For completeness and proper reference, we put the last modified version of the wave function with coulombic part as parent as

$$\psi_{rel+conf}(r) = \frac{N'}{\sqrt{\pi a_0^3}} e^{\frac{-r}{a_0}} \left( C' - \frac{\mu b a_0 r^2}{2} \right) \left( \frac{r}{a_0} \right)^{-\epsilon},\quad (8)$$

where $N'$ is the normalization constant

$$C' = 1 + cA_0\sqrt{\pi a_0^3}\quad (9)$$

$$\mu = \frac{m_i m_j}{m_i + m_j}\quad (10)$$

$$a_0 = \left( \frac{4}{3} \mu \alpha_s \right)^{-1}\quad (11)$$

$$\epsilon = 1 - \sqrt{1 - \left( \frac{4}{3} \alpha_s \right)^2}.\quad (12)$$

The QCD potential [10] is taken as

$$V(r) = -\frac{4}{3r} \alpha_s + br + c.\quad (13)$$

Here $A_0$ is the undetermined factor appearing in the series solution of the Schrödinger equation.

## 4. Leptonic decay constant $f_p$ for different $\Lambda$ scale

The wavefunction in momentum space can be obtained by using the Fourier transform as [13]

$$\psi(p) = \frac{1}{(2\pi\hbar c)^{3/2}} \int d^3r \exp^{-\frac{ipr}{\hbar c}} \psi(r).\quad (14)$$

For $l = 0$, in natural unit,

$$\psi(p) = \sqrt{\frac{2}{\pi p^2}} \int dr \sin(pr)\psi(r).\quad (15)$$

$$\psi_p(p) = \frac{N\sqrt{2}(2-\epsilon)\Gamma(2-\epsilon)}{\pi(1 + a_0^2 p^2)^{\frac{3-\epsilon}{2}}} \left[ C' - \frac{(4-\epsilon)(3-\epsilon)\mu b a_0^3}{2(1 + a_0^2 p^2)} \right]\quad (16)$$

The Decay constant with relativistic correction can be expressed through the meson wave function $\psi(p)$ in the momentum space as [1, 3],

$$f_p = \sqrt{\frac{12}{M_p}} \int \frac{d^3 p}{(2\pi)^3} \left( \frac{E_q + m_q}{2E_q} \right)^{1/2} \left( \frac{E_{\bar{q}} + m_{\bar{q}}}{2E_{\bar{q}}} \right)^{1/2}$$
$$\left( 1 + \lambda_p \frac{p^2}{(E_q + m_q)(E_{\bar{q}} + m_{\bar{q}})} \psi_p(p) \right)\quad (17)$$

with $\lambda_p = -1$ for pseodoscalar mesons and $E_q = \sqrt{p^2 + m_p^2}$.

The strong running coupling constant $\alpha_s$ in the model is considered to be related to the quark mass parameter and an energy scale $\Lambda$ in a specific prescription [11] as

$$\alpha_s(\mu^2) = \frac{4\pi}{\left(11 - \frac{2n_f}{3}\right)\ln\left(\frac{\mu_r^2 + M_B^2}{\Lambda^2}\right)}, \tag{18}$$

where $M_B$ is the background mass, $n_f$ is the number of flavor and $\mu_r = \frac{2m_i m_j}{m_i + m_j}$ is related to quark mass. With this prescription and using the input parameters $b = 0.183$ GeV$^2$, $m_b = 4.79$ GeV and $m_c = 1.55$ GeV, we compute the value of $f_{B_c}$ for different values of $\Lambda$ and tabulate in Tables 1 and 2.

From Table 1, we see that the decay constant gives a comparable result with the other theoretical data for $\Lambda \leqslant 0.150$ GeV.

## 5. The form factors in the semileptonic decay

### 5.1. Slope and curvature of Isgur–Wise function

The Isgur–Wise function [17] is an universal function of velocity transfer which is normalized to unity at zero recoil. In an explicit form, this function can be written as [18]

$$\xi(\omega) = 1 - \rho^2(\omega - 1) + C(\omega - 1)^2 + \ldots, \tag{19}$$

where

$$\omega = v.v' = \frac{(m_{B_c}^2 + m_F^2 - q^2)}{2m_{B_c}m_F}. \tag{20}$$

The quantity $\rho^2$ is the slope of Isgur–Wise function at $\omega = 1$ and $C$ is curvature:

$$\rho^2 = \frac{\partial \xi}{\partial \omega}\Big|_{\omega=1}, \tag{21}$$

$$C = \frac{1}{2}\frac{\partial^2 \xi}{\partial \omega^2}\Big|_{\omega=1}. \tag{22}$$

For the heavy-light mesons, the Isgur–Wise function can also be written as [18]

**Table 1** Decay constants for leptonic decay of $B_c$ with different $\Lambda$

| $\Lambda$ (GeV) | 0.100 | 0.150 | 0.200 | 0.410 |
|---|---|---|---|---|
| $f_{B_c}$ | 0.338 | 0.625 | 0.917 | 1.328 |

**Table 2** Comparison of Decay constant of $B_c$ with others values

| Meson | Decay Constant $f_{B_c}$ (GeV) |
|---|---|
| $B_c$ | 0.338 [Our Work with $\Lambda = 0.100$ GeV] |
| | 0.625 [Our Work with $\Lambda = 0.150$ GeV] |
| | 0.315 [14] |
| | 0.311 [15] |
| | 0.607 [16] |

$$\xi(\omega) = \int_0^\infty 4\pi r^2 |\psi(r)|^2 \cos prdr \tag{23}$$

with

$$p^2 = 2\mu^2(\omega - 1). \tag{24}$$

Now from equations (21) and (22), we compute the slope and curvature of the Isgur–Wise function for different scales of $\Lambda$ as is done for leptonic decay (Table 3).

From the above table, it is realized that the results would be justified for semileptonic decay if we choose the values of $\rho^2$ and $C$ for $\Lambda \geqslant 0.410$ GeV.

### 5.2. Form factors and Isgur–Wise function

In the case of the final states in the decays $B_c \rightarrow \eta_c l\nu$, $B_c \rightarrow B_s l\nu$, and $B_c \rightarrow Bl\nu$ correspond to $J = 0$, as the matrix element of any axial current $A^\mu$ between the two pseodoscalar mesons vanishes, only vector current $V^\mu$ contributes. Unlike in the case of electromagnetic current of the charged pions, here the vector current $V^\mu = \bar{c}\gamma^\mu b$ is not conserved as $q_\mu V^\mu \propto (m_b - m) \neq 0$. So the matrix element of the hadronic current, $V^\mu$ between the two $J^P = 0^-$ mesons is expressed in terms of two form factors $f_\pm(q^2)$ as

$$\langle F(p')|V^\mu|B_c(p)\rangle = f_+(q^2)(p + p') + f_-(q^2)(p - p'), \tag{25}$$

where $q = p - p' = k_1 + k_2$ is the four-momentum transfer, $F = \eta_c, B_s$ and $B$ and $f_+(q^2)$ and $f_-(q^2)$ are the dimensionless weak transition form factors corresponding to $B_c \rightarrow F$, which are functions of the invariant. Here it varies within the range

**Table 3** Slope and curvature of Isgur–Wise function for different $\Lambda$

| $\Lambda$ (GeV) | 0.100 | 0.150 | 0.200 | 0.410 |
|---|---|---|---|---|
| $\rho^2$ | 116.9 | 61.69 | 33.48 | 5.77 |
| $C$ | 5353.4 | 2157 | 908 | 36.78 |

$$m^2_l \leqslant q^2 \leqslant (m_{B_c} - m_F)^2 = q^2_{max}. \tag{26}$$

The transition between the pseudoscalar $B_c$ and the vector mesons depends on four independent form factors as $V(q^2), A_2(q^2), A_1(q^1)$ and $A_0(q^2)$.

These form factors are rigorously calculated with the help of Isgur–Wise function $\xi(\omega)$ as shown in our previous paper [19],

$$f_\pm(q^2) = \xi(\omega)\frac{m_{B_c} \pm m_F}{2\sqrt{m_{B_c} m_F}} \tag{27}$$

$$V(q^2) = A_2(q^2) = A_0(q^2) = \frac{(M_{B_c} + M_{F^*})^2}{4M_{B_c} M_{F^*}}\xi(\omega) \tag{28}$$

$$A_1(q^2) = V(q^2)\left(1 - \frac{q^2}{(M_{B_c} + M_{F^*})^2}\right), \tag{29}$$

The natural expansion parameters for $B_c$ decays to charmonium are the active heavy $b$ and $c$ quark masses as well as the spectator $c$ quark mass. We carry such an expansion up to the second order in the ratios of the relative quark momentum $p$ and binding energy to the heavy quark masses $m_{b,c}$. The numerical results of the scalar and vector form factors are tabulated in Tables 4 and 5.

### 5.3. Decay width

The semileptonic decay width for $B_c \to \eta ev$ for the massless leptons is given by

$$\Gamma(B_c \to \eta ev) = \int \frac{G_F^2}{48\pi^3} V_{cb}^2 M_{\eta_c}^3 (\omega^2 - 1)^{3/2} (M_{B_c} + M_{\eta_c})^2$$
$$\xi^2(\omega)d\omega, \tag{30}$$

where $\omega$ lies in the range $1 \leqslant \omega \leqslant \frac{M_{B_c}^2 + M_{\eta_c}^2 - m_l^2}{2M_{B_c} M_{\eta_c}}$.

The Decay width is calculated for $\Lambda = 410$ MeV, $M_{B_c} = 6.227$ GeV, $M_{\eta_c} = 2.983$ GeV, $b = 0.183, c = -0.37$ GeV, $n_f = 4$ is found to be $\Gamma = 6.38166 \times 10^{-10} V_{cb}^2$.

**Table 4** Form factors of weak decay of $B_c$ into scalars

| Transitions | – | $f_+(0)$ | $f_-(0)$ | $f_+(q^2_{max})$ | $f_+(q^2_{max})$ |
|---|---|---|---|---|---|
| $B_c \to \eta_c ev$ | Our work | 2.1289 | 0.757 | 1.07011 | 0.3809 |
| | [20] | 0.47 | 0.47 | 1.07 | 0.92 |
| $B_c \to B_s ev$ | Our work | 0.9441 | 0.0738 | 1.003 | 0.0784 |
| | [20] | 0.50 | 0.50 | 0.99 | 0.99 |
| $B_c \to Bev$ | Our work | 0.9319 | 0.0804 | 1.003 | 0.08668 |
| | [20] | 0.39 | 0.39 | 0.96 | 0.80 |

**Table 5** Form factors of weak decay of $B_c$ into vectors

| Transitions | – | $V(0)$ | $A_1(q^2_{max})$ | $V(q^2_{max})$ |
|---|---|---|---|---|
| $B_c \to J/\psi ev$ | Our work | 2.248 | −0.186 | 1.130 |
| | [20] | 0.49 | 0.88 | 1.34 |
| $B_c \to B_s^* ev$ | Our work | 0.9463 | 1.0014 | 1.0054 |
| | [20] | 3.44 | 0.76 | 6.25 |
| $B_c \to B^* ev$ | Our work | 0.934 | 1.00064 | 1.00678 |
| | [20] | 3.94 | 0.72 | 8.91 |

Taking the mean values of lifetimes from PDG 2012 [21], $\tau = 0.507ps$ and also using the average value of $V_{cb} = 0.041$ we obtain the Brancing ratio $B = 1.3 \times 10^{-5}$ by using the relation $B = \Gamma \times \tau$. This result can be compared to some extent with the experimental data of $B = 5.2 \times 10^{-5}$ from Particle Data Group [22]. However, this result has a deviation from the theoretical evaluation of reference [7] for the channel $B_c \to \eta l\bar{v}$ which is $1.85 \times 10^{-7}$ in the LO and $2.33 \times 10^{-7}$ in the NLO.

## 6. Conclusions

In this manuscript, we have studied the leptonic and semileptonic decay of $B_c$ meson for different energy scales. For leptonic decay, the results are comparable to other available data with a scale of $\Lambda \leqslant 0.150$ GeV, whereas in case of semileptonic decay the results show an agreement for $\Lambda \geqslant 0.410$ GeV. Thus, a two-scale picture is evident for any potential model to study the leptonic and semileptonic decay of mesons.